\documentclass[prb,twocolumn,superscriptaddress,showpacs]{revtex4}
\usepackage{latexsym,graphicx,amsmath,amsfonts}

\def\vec#1{{\bf #1}}

\def\ket#1{| #1 \rangle}

\def\bra#1{\langle #1 |}

\def\proj#1{\ket{#1}\bra{#1}}

\begin{document}
\title{Robust Charge-based Qubit Encoding}
\author{Daniel~K.~L.~\surname{Oi}}
\email{D.K.L.Oi@damtp.cam.ac.uk}
\affiliation{Department of Applied Mathematics and Theoretical Physics, University of
    Cambridge, Wilberforce Road, Cambridge CB3 0WA, UK}
\author{Sonia~G.~\surname{Schirmer}}
\affiliation{Department of Applied Mathematics and Theoretical Physics, University of
    Cambridge, Wilberforce Road, Cambridge CB3 0WA, UK}
\author{Andrew~D.~\surname{Greentree}}
\affiliation{Centre for Quantum Computer Technology, University of Melbourne,
  Victoria, Australia}
\author{Tom~M.~\surname{Stace}}
\affiliation{Department of Applied Mathematics and Theoretical Physics, University of
    Cambridge, Wilberforce Road, Cambridge CB3 0WA, UK}
\date{\today}

\begin{abstract}
  We propose a simple encoding of charge-based quantum dot qubits which
  protects against fluctuating electric fields by charge symmetry of the
  logical states. We analyze the reduction of coupling to noise due to nearby
  charge traps and present single qubit gates. The relative advantage of the
  encoding increases with lower charge trap density.
\end{abstract}


\maketitle

\section{Introduction}

Quantum computation faces considerable hurdles, one of the most serious being
engineering physical systems performing coherent operations without the
deleterious effects of decoherence~\cite{Zurek1991}, particularly in the solid
state. However by isolation and manipulation of states of quantum dot (QD)
structures~\cite{Divincenzo2000}, it may be possible to perform many unitary
operations within the dephasing time, a pre-requisite for quantum error
correction (QEC) by means of Calderbank-Shor-Steane codes~\cite{Steane1996}.

Underlying logical QEC, a complementary strategy is to use Hilbert
subspaces which couple least to noise processes, decoherence free
subspaces (DFS)~\cite{Zanardi1997,Lidar1998,Zanardi1998,FIGDB2004}.
Practical quantum computing will undoubtedly use elements of both.
Charge-based QD quantum
computing~\cite{EkertJozsa,HFCJH2003,SGO2003,Petta2004,Gorman2005,Buehler2005}
is a prime candidate for DFS encoding as electric field coupling is
a major source of decoherence~\cite{bib:Hollenberg2003,BM2003}.
Here, we present an architecture incorporating charge symmetry of
the logical states to protect against electromagnetic fluctuations,
analyze its resistance to charge trap noise and present single-qubit
gates. Coupling to charge trap noise and decoherence is suppressed
by several orders of magnitude compared to a conventional charge
qubit, depending on charge trap density.  Alternatives to the
passive control implied by DFS encoding include active control
sequences, such as Bang-Bang control \cite{Viola1998,Fraval2004}.

 \begin{figure}
 \includegraphics[width=0.45\textwidth]{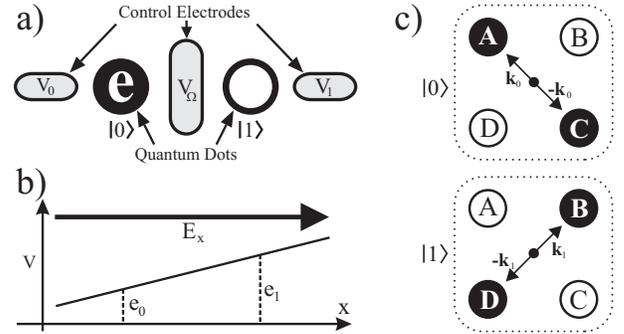}
 \caption{a) 2QD Dipole Charge Qubit. Logical states, $\{\ket{0},\ket{1}\}$,
   are defined by an excess electron on the left or right dot respectively.
   Symmetry electrodes ($V_{0,1}$) control $\sigma_z$ rotations, and a barrier
   electrode ($V_\Omega$) controls $\sigma_x$. An SET (omitted) measures in the
   logical basis.  b) Temporally and spatially varying potential due to external
   field. c) 2-Electron 4QD Quadrupole Qubit with $\ket{0}=a^\dagger
   c^\dagger\ket{\text{vac}},\ \ket{1}=b^\dagger d^\dagger\ket{\text{vac}}$.}
 \label{fig:chargequbit}
 \end{figure}

In a typical charge-based QD qubit (Fig.~\ref{fig:chargequbit}a) the position
of an excess electron defines the logical states. Ideally, the logical states
of the system should be eigenstates of the system Hamiltonian when the system
is idle, i.e. tunnelling should be suppressed on practical time-scales by
$V_\Omega$. Furthermore, we assume that the system can be tuned, via $V_{0,1}$,
such that the logical states are degenerate, hence (known) relative dynamical
phases can be neglected.

\section{Classical Noise}

Fluctuations of the electromagnetic environment superimpose inhomogeneities on
the potential seen by the charge states. An electric field component along the
axis of the qubit will cause a sloping potential~(Fig.~\ref{fig:chargequbit}b),
inducing for each state a different dynamic phase,
\begin{equation}
\ket{j(t)}=e^{-\frac{i q}{\hbar}\int_0^t \epsilon_j(t') dt'}\ket{j(0)},\ j=0,1
\label{eq:phasefluct}
\end{equation}
where $\epsilon_j$ is the on-site energy fluctuation and $q$ the electron
charge. Fluctuations drive superpositions
$\ket{\psi}=\alpha\ket{0}+\beta\ket{1}$ to mixed states,
$\proj{\psi}\mapsto|\alpha|^2\proj{0}+|\beta|^2\proj{1}$.  Furthermore,
electrodes operating on nearby qubits will look like noise, i.e. it may only
be practical to actively compensate for operations on nearest neighbors, but
not those further away which may also cause unwanted perturbations.


We generalize Eq.~(\ref{eq:phasefluct}) to multi-electron configurations,
encoding logical states in many-particle states whose geometry protects against
decoherence~(Fig.~\ref{fig:chargequbit}c)~\cite{TOALBS1999,TothLent2001}. Two
excess electrons in diagonally opposite dots define the logical states.  Single
square QDs in the limit of large dot size should display similar
dynamics~\cite{CreffieldPlatero2002,JFTS2002}. The 4QD arrangement has also
been considered for Coherent Quantum Cellular
Automata~\cite{Jared2004,GSCRL2003}, and for scalable
qubits~\cite{TothLent2001}. Measurement in the logical basis can be achieved by
a single electron transistor (SET) adjacent to one of the dots in each
qubit~\cite{bib:Hollenberg2003}, or by using multiple SETs in a correlated
mode~\cite{BRBHDC2003}. Qutrits or higher dimensional systems may also be
considered, e.g. a qutrit encoded as two electrons in a three dimensional 6QD
octohedral structure.

An external electric field induces phase shifts, as in
Eq.~(\ref{eq:phasefluct}), where the energies to first order are
\begin{eqnarray*}
\epsilon_0(t)&=&\epsilon_A+\epsilon_C=
2\bar{V}+\vec{k_0}\cdot\vec{E}+(-\vec{k_0})\cdot\vec{E}=2\bar{V}\\
\epsilon_1(t)&=&\epsilon_B+\epsilon_D=
2\bar{V}+\vec{k_1}\cdot\vec{E}+(-\vec{k_1})\cdot\vec{E}=2\bar{V},
\end{eqnarray*}
where $\bar{V}$ is the potential at the common centroid and
$\epsilon_{A,B,C,D}$ are the on-site energy fluctuations of the respective QDs.
The symmetrical distributions of charge ensure that each logical state acquires
the same dynamic phase due to the external potential gradient. Thus, an
initial superposition acquires an \emph{overall} dynamic phase which is
unobservable.

\section{Charge Trap Noise}

Though linear spatially varying potentials have no dephasing effect
on the 4QD qubit, charge trap fluctuators~\cite{BRBHDC2003} may pose
a
problem~\cite{ItakuraTokura2003,Schultz83,Mueller98,McCamey2005,Ang2005}.
An occupied charge trap has a $\sim1/r$ potential, which perturbs
the degeneracy of the DFS states.  In principle the charge trap
density can be made arbitrarily low but a few charge traps may be
unavoidable in practice~\footnote{For thick $\text{SiO}_2$ layers on
Si, defect densities of $\sim 10^8 \mathrm{cm}^{-2}$ have been
achieved~\cite{Schultz83}. Trap densities for thin oxide layers of
order $\sim 10^{11}\mathrm{cm}^{-2}$ have been reported in
Refs.~\onlinecite{Mueller98,McCamey2005}}, and charge trap noise may
be a significant source of decoherence.

\begin{figure}
  \includegraphics[width=0.45\textwidth]{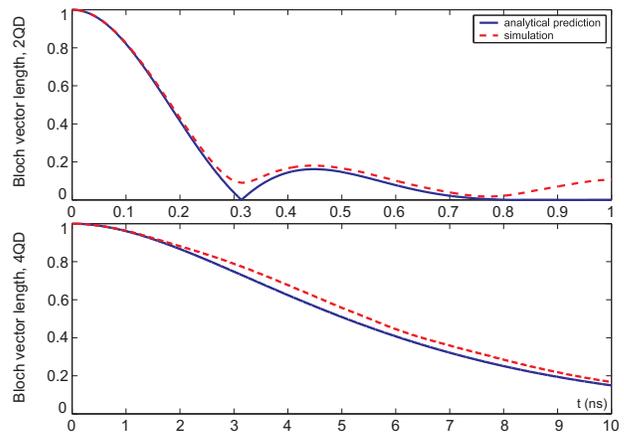}
  \caption{Qubit Coherence Decay. Initial state
    $(\ket{0}+\ket{1})/\sqrt{2}$ coupled to 100 charge traps.
    $k_\text{eff}^{(2)}=12.46\times10^9\hbar/s$ and
    $k_\text{eff}^{(4)}=0.60\times10^9\hbar/s$. A total of 200 quantum
    trajectories were simulated and averaged. For all simulations, the 2QD
    qubit was $20$nm long and the 4QD qubit was a $20$nm square (e.g. P donors
    in Si, see Ref.~\onlinecite{bib:Hollenberg2003}).  Both were located $20$nm below
    the layer in which the charge traps were located.  Charge trap transition
    rate was $2\times10^8$Hz.}
  \label{fig:typicaldecay}
  \end{figure}

\begin{figure}
   \includegraphics[width=0.45\textwidth]{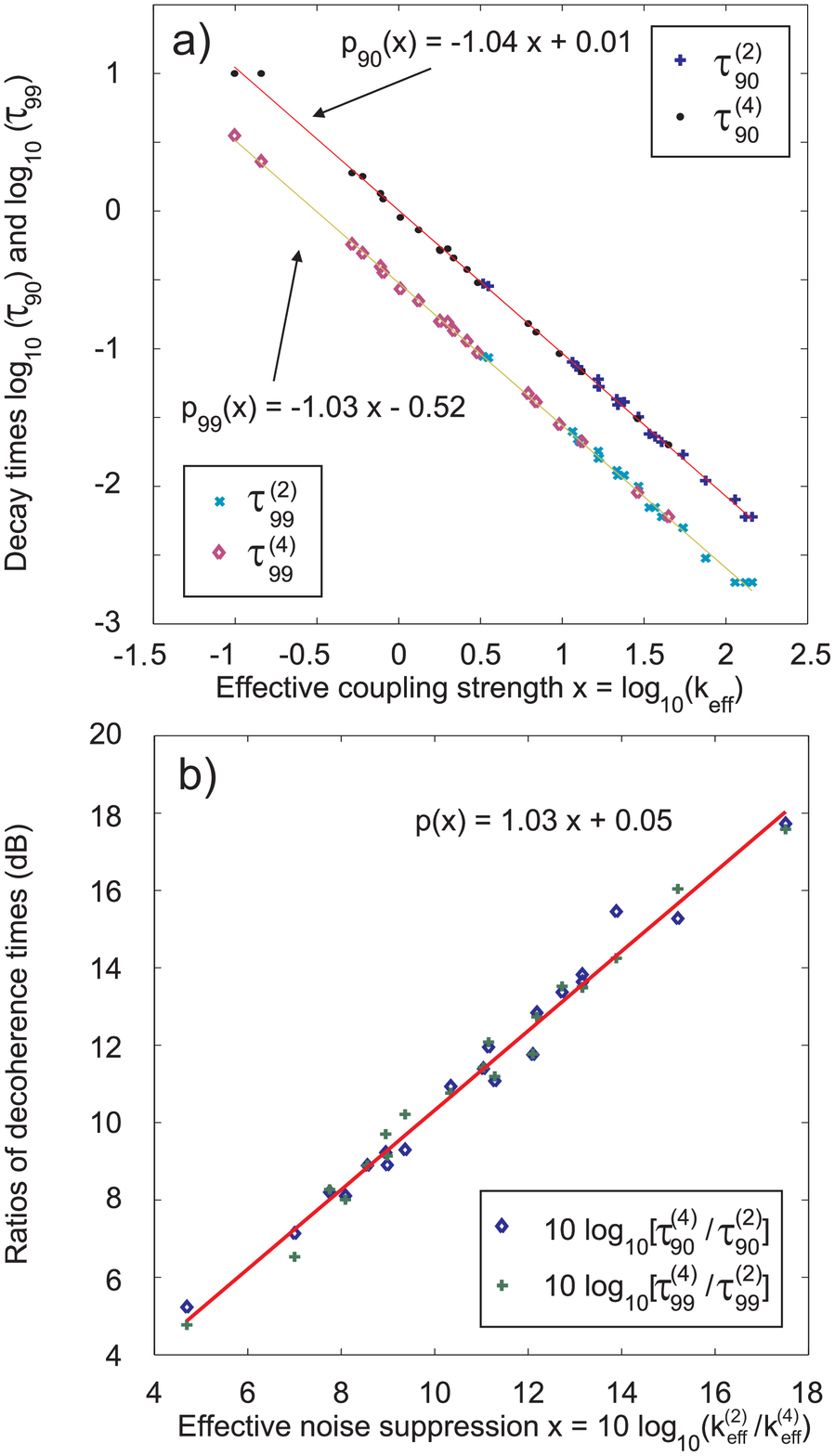}
 \caption{Decoherence of 2QD and 4QD qubits.
   (a) The decoherence decay times are inversely related to the effective coupling
   to fluctuators and show the same dependence for the 2QD and 4QD qubits.
   (b) The ratio of the short-term coherence times for the 2QD and 4QD encodings
   are inversely proportional to the ratio of the effective coupling constants.
   Each point represents the average of 200 quantum trajectories of a qubit coupled
   to 100 randomly distributed fluctuators.}
 \label{fig:decaytimes}
 \end{figure}

To understand the effect of charge trap noise, consider a single charge trap coupled
to the qubit via the Hamiltonian $H=k\xi(t)\sigma_z/2$, where $\xi=\pm1$ is a Poisson
process of rate $\lambda$, and $k$ is the coupling~\cite{GAS2003,ML2004}. Averaging
over noise processes leads to a decay of the coherence of the qubit density operator,
\begin{eqnarray*}
\langle\rho_{01}(t)\rangle_\xi&=&\rho_{01}(0)\left\langle e^{-ik\int_0^t
  \xi(t')dt'}\right\rangle_{\xi}\nonumber\\
&=&\rho_{01}(0)e^{-t\lambda}\left[\cos\omega t+\frac{\lambda}{\omega}\sin\omega t\right]
\label{eqn:singlesolution}
\end{eqnarray*}
where $\omega=\sqrt{k^2-\lambda^2}$. For many independent fluctuators with different
rates $\lambda_j$ and couplings $k_j$, the coherence decays in a non-Markovian manner
(Fig.~\ref{fig:typicaldecay}),
\begin{equation}
\langle\rho_{01}(t)\rangle_\xi=\rho_{01}(0)e^{-t\sum_j
  \lambda_j}\prod_j\left[\cos\omega_j t+\frac{\lambda_j}{\omega_j}\sin\omega_j t\right].
\label{eqn:manyfluct}
\end{equation}
A Taylor expansion of the solution~(\ref{eqn:manyfluct}) about $t=0$ shows that the
initial decay is parabolic
\begin{equation}
  \langle\rho_{01}\rangle_\xi/\rho_{01}(0) \approx1-t^2/2\sum_j k_j^2+O(t^3),\quad t\ll1,
\end{equation}
independent of $\lambda_j$, and depends only on the effective coupling of the encoded
qubits to the charge traps, $k_\text{eff}^2=\sum_j k_j^2$ (Fig.~\ref{fig:decaytimes}a).
Therefore, the short-term behavior will be dominated by the fluctuator that couples
most strongly to the qubit, while the others mainly dampen further oscillations of
the coherence vector (Fig.~\ref{fig:typicaldecay}).  Furthermore, the time it takes
for the coherence to decay from $1$ (maximal coherence) to $p$ for $p$ close to $1$,
which is of crucial importantance in quantum information processing, is inversely
proportional to the effective coupling strength $k_\text{eff}$, $\tau_p=\sqrt{2(1-p)}
k_\text{eff}^{-1}$, and we have
\begin{equation}
   \frac{\tau_p^{(4)}}{\tau_p^{(2)}} = \frac{k_\text{eff}^{(2)}}{k_\text{eff}^{(4)}},
\end{equation}
where $k_\text{eff}^{(2)}$ and $k_\text{eff}^{(4)}$ is the effective coupling strength
for the two-dot and four-dot encoding respectively (Fig.~\ref{fig:decaytimes} b).
Thus, the ratio of the effective coupling strengths is a good measure for the
superiority of the 4QD encoding versus the 2QD encoding---the former will be better
provided that $k_\text{eff}^{(2)}/k_\text{eff}^{(4)}>1$, and the larger the ratio the
greater the improvement.

\begin{figure}
 \includegraphics[width=0.45\textwidth]{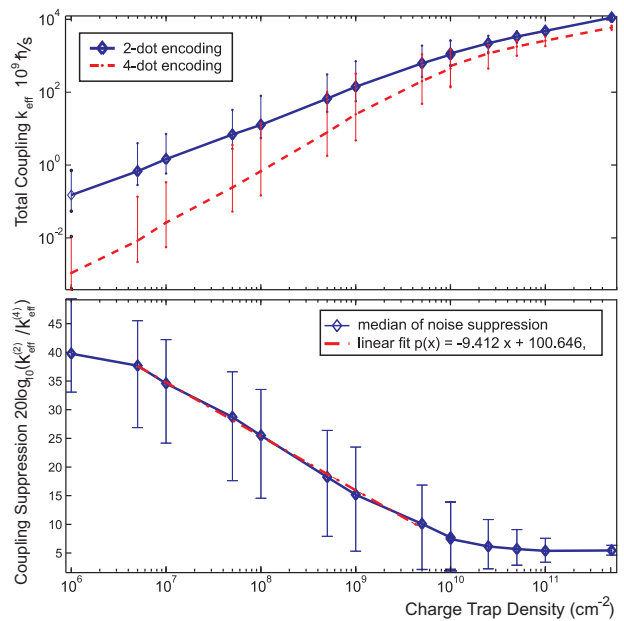}
 \caption{Coupling and Noise vs Charge Trap Density. Noise suppression of the
   4QD vs 2QD qubits was calculated for 100 random charge trap distributions per
   density, which was then averaged. Coupling strength is in units of
   $10^9\hbar/s$, bars indicate the $10\%-90\%$ ranges. At high densities, the mean
   charge trap spacing is comparable to the size of the qubit, leading to
   saturation effects.}
 \label{fig:trapdensity}
 \end{figure}

The 2QD and 4QD qubits couple differently to charge traps, $k_{j}^{(2)}\propto r_j^{-2}$,
and $\ k_{j}^{(4)}\propto r_j^{-3}$ respectively, where $r_j$ is the distance between the
qubit and each charge trap.  The 4QD qubit has thus effectively a smaller ``horizon''
than the 2QD qubit.  Hence, generally a charge trap would have to be situated closer
to the 4QD qubit than a 2QD qubit to induce the same decoherence.  Since the noise on
the qubit is generally dominated by the closest fluctuator, whose typical distance is
inversely proportional to defect density, the average relative effectiveness of the
encoding is therefore expected to increase with decreasing charge trap density, which
is confirmed by computer simultations (Fig.~\ref{fig:trapdensity}).

\section{Robustness of Encoding}

The results in the previous section show that the 4QD encoding can substantially
increase short-term coherence times for an ideal geometry.  However, any physical
implementation is likely to deviate from the perfect symmetry of the ideal quantum
dot structure.  The scheme's sensitivity to such deviations is thus an important
practical consideration.

Non-ideal geometry, e.g. due to imperfect QD placement, will introduce a dipole
moment, spoiling decoupling from external fields and reducing robustness to charge
trap noise.  However, as the magnitude of this dipole is comparable to, and linear
in the displacement, and given that fabricational precision should be at least a
fraction of QD spacing for QIP purposes~\footnote{Atomic-scale placement ($\sim 1$
nm accuracy) of P donors in Si has been demonstrated~\cite{Schofield2003}.}, the
extra dipole for the 4QD qubit should be much smaller than for a 2QD qubit, hence
the encoding should still offer a noticeable advantage.

\begin{figure}
   \includegraphics[width=0.45\textwidth]{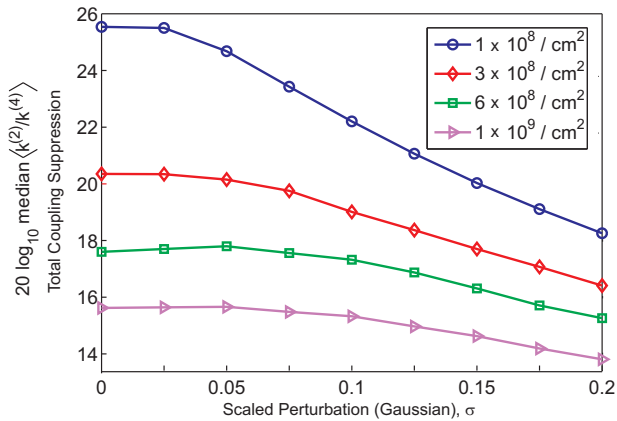}
\caption{Decoupling vs Placement Error. We simultaneously and independently
  purturbed the positions of all quantum dots by a Gaussian displacement with
  standard deviation $\sigma$ (expressed as a fraction of the array
  side-length). This was repeated 1000 times for each of 500 fluctuator
  distributions. We plot of median w.r.t fluctuator distributions of the mean
  over the perturbations of the decoupling ratios as a function of error
  magnitude for different effective fluctuator densities. Even with a 10\%
  displacement error (i.e. $\sigma = 0.1$), the 4QD qubit is still effective.}
 \label{fig:perturb}
 \end{figure}

To quantify the effect of asymmetry due to placement errors in the quantum dots,
we performed extensive simulations computing the effective couplings for various
randomly perturbed 2QD and 4QD architectures for different charge trap densities
and a wide range of charge distributions.  The simulations show that for reasonable
errors ($\sim 10\%$ placement error), the efficiency of the scheme is only modestly
affected over a wide range of fluctuator densities (Fig.~\ref{fig:perturb}).

\section{Quantum Gates}

We now consider implementing a universal set of quantum gates,
$\{\sigma_z^\phi,\sigma_x^{\pi/2},c-\sigma_z^\pi\}$. Ideally, we
would like all states involved during gate operations to belong to the DFS.
This suggests an adiabatic holonomic control
scheme~\cite{ZR1999,USB1999,DCZ2001}. However, the requirement of additional
quantum dots for generating holonomies, charge symmetry constraints on
auxiliary dot positioning, and the complexity of pulse sequences all offset
possible advantages of holonomic control. Alternatively, rapidly modulating
tunneling between dots can implement the required logical gates quickly. If the
gate time is short enough, transient population in non-DFS states, should have
minimal coupling to electric field fluctuations.  Intra-dot charging should
effectively suppress double occupation, which can be further enhanced by
ensuring all spins are parallel so that each orbital cannot have more than one
electron.


We describe the four dot system as a two-electron, four-site Hubbard
model. The electronic creation operators are defined $a^{\dagger},
b^{\dagger}, c^{\dagger}, d^{\dagger}$ for dots $A,B,C,D$
respectively, labelled in clockwise fashion
(Fig.~\ref{fig:chargequbit}c). Firstly, the phase gate (a rotation
of the Bloch sphere by angle $\phi$ around the z-axis)
$\sigma_z^\phi=\text{diag}(1,e^{i\phi})$ is achieved by biasing one
pair of diagonally opposite quantum dots with respect to the other,
\begin{equation*}
\phi=\frac{2e}{\hbar}\int_0^t\left[V_0(t')-V_1(t')\right]dt',
\end{equation*}
where $V_0,V_1$ are the on-site potentials of the quantum dots
defining the $\ket{0}\equiv a^\dagger c^\dagger\ket{\text{vac}}$ and
$\ket{1}\equiv b^\dagger d^\dagger\ket{\text{vac}}$ states
respectively.

Next, the $\sigma_x^{\pi/2}$ gate requires inter-dot tunnelling. We allow
tunnelling between dots $A\leftrightarrow D$ and $B\leftrightarrow C$.
Allowing $A\leftrightarrow B$ and $C\leftrightarrow D$ tunneling as well leads
to similar dynamics but at the expense of extra control electrodes and more
non-DFS states involved.  With vertical tunneling only, the available state
space is spanned by four states, $\{\ket{0},\ket{1},\ket{\varepsilon_0}=a^\dagger
b^\dagger\ket{\text{vac}},\ket{\varepsilon_1}=c^\dagger
d^\dagger\ket{\text{vac}}\}$. Hence the Hamiltonian with no tunneling is, $
H_{0}^{V}=\text{diag}(0,0,\delta,\delta)
$ in the above basis, and where we have taken the (degenerate) ground state
energy to be $0$ and $\delta$ is the energy of the non-diagonal states
$\{\ket{\varepsilon_0},\ket{\varepsilon_1}\}$ due to Coulomb repulsion of the
two electrons. We now switch on equal tunneling with rate $\Omega$ in the
vertical direction,
\begin{equation*}
H_\text{tunnel}^{V}=\left(
\begin{array}{cccc}
0 & 0 & \Omega & \Omega \\
0 & 0 & \Omega & \Omega \\
\Omega & \Omega & 0 & 0 \\
\Omega & \Omega & 0 & 0
\end{array}
\right).
\end{equation*}
For convenience, we normalize $\delta=1$, and scale $\Omega$ relative
to this. The eigenstates of $H_\text{tot}=H_{0}+H_\text{tunnel}$ are
\begin{eqnarray*}
\ket{\psi_1}&=&\ket{0}-\ket{1}, \\
\ket{\psi_2}&=&\ket{\varepsilon_0}-\ket{\varepsilon_1},\nonumber\\
\ket{\psi_3}&=&
\frac{4\Omega(\ket{0}+\ket{1})}{\sqrt{1+16\Omega^2}+1}
+\ket{\varepsilon_0}+\ket{\varepsilon_1},\nonumber\\
\ket{\psi_4}&=&
\frac{4\Omega(\ket{0}+\ket{1})}{\sqrt{1+16\Omega^2}-1}
-\ket{\varepsilon_0}-\ket{\varepsilon_1},
\label{eq:eigs}
\end{eqnarray*}
where the eigen-energies are
$E_1=0,E_2=1,E_3=(1+\sqrt{1+16\Omega^2})/2,E_4=(1-\sqrt{1+16\Omega^2})/2$.
Tunnelling between dots mixes the states so that $\{\ket{0},\ket{1}\}$ are no
longer eigenstates of $H_\text{tot}$. Transitions between $\ket{0}$ and
$\ket{1}$ cannot occur directly but only via transient occupation of the
non-DFS states.

In order to achieve a $\pi/2$ rotation around an axis
$(\cos\gamma,\sin\gamma,0)$ lying on the equator of the Bloch sphere (which is
equiavalent to a $\sigma_x^{\pi/2}$ gate up to $\sigma_z^\gamma$ rotations)
$\ket{0}\mapsto \ket{0} + e^{i\gamma}\ket{1}, \quad \ket{1}\mapsto
\ket{0}-e^{i\gamma}\ket{1}$, we require $E_3$ and $E_4$ to be rational.  This
leads to the conditions, $4\Omega=\sqrt{(n/m)^2-1}$ where
$\{n/2,m\}\subset\mathbb{Z}^{+}$, $n>m$ and $\text{gcd}(n,m)=1$.  These
requirements derive from the fact that the amplitudes of the $\ket{0}$ and
$\ket{1}$ should be equal in magnitude when the amplitudes of the non-DFS
states are zero, leading to $jm/n=1/2+k,\{j,k\}\subset\mathbb{Z}^+$, and the
gate time $t_f=2jm\pi/n$. When $t_f=\pi m$, we achieve the operation with
$\gamma=\pi(n-m)/2$. If $t_f=2\pi m$, we perform a logical NOT
($\ket{0}\leftrightarrow\ket{1}$).

\begin{figure}
 \includegraphics[width=0.45\textwidth]{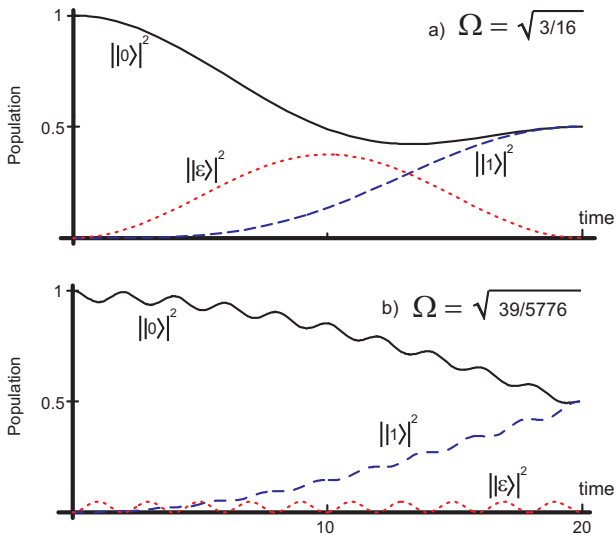}
 \caption{Populations During $\sigma_x^{\pi/2}$-Gate vertical tunneling.
   An initial state $\ket{0}$ is transformed into
   $(\ket{0}+i\ket{1})/\sqrt{2}$ with transient
   population in non-DFS subspace $\ket{\varepsilon}$. a) $\Omega$ large: $n=2,\
   m=1$, b) $\Omega$ small: $n=20,\ m=19$.}
 \label{fig:sigmax}
 \end{figure}

The minimum gate time for a $\pi/2$-gate is $t_f=\pi$ when $2\le n$ even and
$m=1$ (Fig.~\ref{fig:sigmax}).  If $m>1$, the time required to implement the
gate increases. Coulomb repulsion favors the diagonal charge configurations but
transient population in the other states will still occur.  For $n/m\rightarrow
1$, the gate time is on the order of $\pi m$ but the maximum transient
population scales as $(n^2-m^2)/n^2$. The integrated population in the non-DFS
states during the total gate time is proportional to $m(n^2-m^2)/n^2$ and thus
using smaller tunneling rates does not improve the overall transient occupation
of non-DFS states.

 \begin{figure}
 \includegraphics[width=0.45\textwidth]{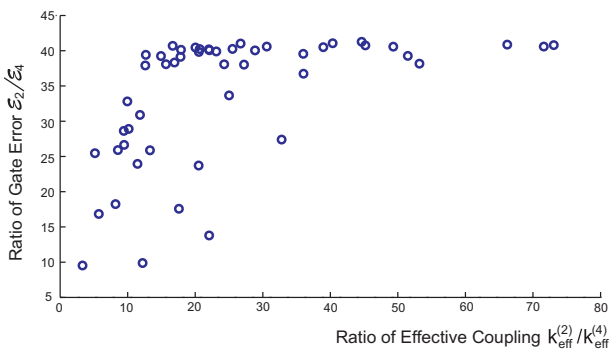}
 \caption{$\sigma_x^{\pi/2}$-Gate Error vs Noise Coupling.  Gate parameters:
   $t_f\approx50\text{ps},\ \Omega_2=\pi/(4t_f),\ \delta=3.84\times10^{12}\hbar/s,
   \ 4\Omega_4=\delta_c\sqrt{(62/61)^2-1}$. Fidelity was calculated from 50
   trajectories per initial state $\{\ket{\pm x},\ket{\pm y},\ket{\pm z}\}$.}
 \label{fig:gatefidelity}
 \end{figure}

The average gate error, $\mathcal{E}=1-\mathcal{F}$ where $\mathcal{F}$ is the
average fidelity~\cite{PC1997,Bowdrey2002}, for different charge trap couplings
(densities) was simulated for 4QD and 2QD qubits. The ratio of the errors,
presented in Fig.~\ref{fig:gatefidelity}, show that despite transient
population in non-DFS states during the operation of the
$\sigma_x^{\pi/2}$-gate, the 4QD configuration still shows a significant
advantage over the 2QD qubit.

A universal two-qubit controlled-phase ($c-\phi$) gate may be implemented as
suggested in earlier work~\cite{SchirmerOiGreentree2004}. A transient
deformation of the charge distribution of adjacent qubits by the use of auxiliary
quantum dots would allow modulation of an effective $\sigma_z\otimes\sigma_z$
interaction.  Charge symmetry could still be maintained during the gate by use
of auxiliary dots.

\section{Conclusion}

We have analyzed the noise suppression of a 2-electron 4QD qubit encoding,
which decouples from linearly varying fields. For nearby charge fluctuators,
the decoupling depends upon the exact distribution, but analytic and numerical
results show considerable enhancement of noise resistance and coherence times,
increasing at low charge trap densities. This advantage persists in the
presence dot placement errors.  We show how to construct single qubit
rotations, and two-qubit gates may be possible via previously proposed schemes
for conventional charge qubits.  Characterization and tuning of the 4QD qubit
should be be similar to that for a conventional 2QD qubit. The architecture
requires only a modest increase in complexity and may also be applied to
systems such as super-conducting charge qubits.

\acknowledgments

SGS, DKLO and TMS acknowledge Fujitsu, Cambridge-MIT Institute, EPSRC (UK), and
EU grants RESQ (IST-2001-37559) and TOPQIP (IST-2001-39215). DKLO is supported by
Sidney Sussex College Cambridge. ADG is supported by the Australian Research
Council, the Australian government, and NSA, ARDA and ARO under
DAAD19-01-1-0653. ADG acknowledges the generosity of Fujitsu whilst visiting
Cambridge. We are grateful to Christina Goldschmidt for useful discussions.


\end{document}